\documentclass[10pt,letterpaper]{article}
\usepackage{opex3}
\usepackage{amssymb}
\usepackage{amsmath}
\usepackage{graphicx}
\usepackage{bm}
\usepackage{cite}

\begin{document}

\title{Polarization-Sensitive Quantum Optical Coherence Tomography: Experiment}

\author{Mark~C.~Booth,$^{\bf 1}$ Bahaa~E.~A.~Saleh,$^{\bf 2,3}$ and Malvin~Carl~Teich$^{\bf 1,2,4,*}$}

\address{$^{1}$Department of Biomedical Engineering, Boston University, Boston, MA 02215, USA\\
$^{2}$Quantum Photonics Laboratory, Department of Electrical \&
Computer Engineering,\\ Boston University, Boston, MA 02215, USA\\
$^{3}$Quantum Photonics Laboratory, College of Optics and Photonics
(CREOL),\\ University of Central Florida, Orlando, FL 32816, USA\\
$^{4}$Department of Electrical Engineering, Columbia University, New
York, NY 10027, USA}

\email{$^{*}$teich@bu.edu}
\homepage{http://people.bu.edu/teich/} 

\hyphenation{wave-guide wave-guides}

\begin{abstract}
Polarization-sensitive quantum optical coherence tomography
(PS-QOCT) makes use of a Type-II twin-photon light source for
carrying out optical sectioning with polarization sensitivity. A BBO
nonlinear optical crystal pumped by a Ti:sapphire psec-pulsed laser
is used to confirm the theoretical underpinnings of this imaging
paradigm. PS-QOCT offers even-order dispersion cancellation with
simultaneous access to the group-velocity dispersion characteristics
of the interstitial medium between the reflecting surfaces of the
sample.
\end{abstract}



\section{Introduction} \label{sec:intro}
\paragraph{Optical coherence tomography (OCT).}
Optical coherence-domain reflectometry (OCDR) is a high-resolution
axial-imaging technique originally developed for optical-fiber and
integrated-optics applications \cite{youngquist87, takada87,
danielson87}. It offers high resolution, typically of the order of
$\mu$m, which is achieved via the use of interferometry and the
broad bandwidth of the optical source \cite{gilgen89}.

With the addition of transverse scanning, this range-finding
technique provides cross-sectional images and is called optical
coherence tomography (OCT) \cite{huang91}. Because OCT has the
merits of being a non-invasive, non-contact, high-resolution, and
rapid procedure, it has revolutionized \emph{in vivo} biological
imaging
\cite{huang91,schmitt99,fercher02,fercher03,drexler04,tomlins05,brezinski06,
zysk07}. It has found widespread use in clinical medicine,
particularly in the specialties of ophthalmology, gastroenterology,
dermatology, laryngology, cardiology, and oncology
\cite{zysk07,drexler08}.

OCT makes use of the second-order coherence properties of a
classical optical source to effectively section a reflective or
scattering sample with a resolution governed by the coherence length
of the source. Sources of short coherence length, with their
attendant broad spectral width, provide high resolution. Examples of
optical sources used in OCT include superluminescent diodes (SLDs),
ultrashort pulsed lasers, wavelength-swept lasers, supercontinuum
sources such as photonic-crystal fibers \cite{drexler08}, and
wavelength-swept amplified-spontaneous-emission (ASE) sources
\cite{eigenwillig09}.

OCT can be carried out in a number of modalities, including
time-domain, Fourier-domain, swept-source, Doppler, spectroscopic,
and polarization-sensitive configurations (PS-OCT) \cite{drexler08}.
This latter modality, in particular, reveals enhanced information
about the birefringence properties of samples endowed with an
organized linear structure \cite{hee92}, such as tissues containing
a high content of collagen or other elastin fibers, including
tendons, muscle, nerve, and cartilage \cite{park08}. Indeed, a
change in birefringence can be indicative of a change in the
functionality, integrity, or viability of biological tissue.

\paragraph{Deleterious effects of dispersion.}
The development of improved broadband optical sources has served to
enhance OCT resolution. At the same time, however, the larger
bandwidth leads to increased sample dispersion that increases the
width of the coherence-envelope of the probe beam, which in turn
results in a loss of axial resolution and fringe visibility. In the
particular case of ophthalmologic imaging, for example, the retinal
structure of interest is located behind the thick dispersive ocular
medium.

To further improve the sensitivity of OCT, it is useful to implement
techniques for handling dispersion. Indeed, a whole host of
techniques for mitigating its presence have been developed, ranging
from the use of dispersion-compensating optical elements to \emph{a
posteriori} numerical methods. The use of many of these techniques,
however, requires an understanding of the character of the
dispersion inherent in the sample so that the appropriate optical
element or numerical algorithm can be used
\cite{fercher02,drexler04}.

\paragraph{Advent of nonclassical light.}
Over the past several decades, a number of nonclassical (quantum)
sources of light have been
developed~\cite{teich88,teich89,teich90,perina94,mandel95} and it is
natural to inquire whether making use of any of these sources might
be advantageous for tomographic imaging. An example of such a
nonclassical source is spontaneous parametric down-conversion (SPDC)
\cite{harris67,magde67,klyshko80}, a nonlinear process that
generates entangled twin beams of light. Type-I and Type-II SPDC
interactions generate light with spectral/spatial and polarization
entanglement, respectively \cite{klyshko80}\cite[chap.~21]{saleh07}.
Such sources, which are broadband, have been utilized to demonstrate
a number of interference effects in physics that cannot be observed
using traditional classical sources of light \cite[as an
example]{larchuk93}.

It turns out that we are indeed able to make use of
frequency-entangled twin photon pairs to construct a system that
provides range measurements analogous to those obtained using
classical OCT, but with additional salutary nonclassical features.
We refer to this scheme as quantum optical coherence tomography
(QOCT).

\paragraph{Quantum optical coherence tomography (QOCT).}
QOCT provides axial imaging by making use of a fourth-order
interferometer incorporating two photodetectors, rather than the
second-order interferometer with a single photodetector, as used for
classical OCT. Type-I QOCT was first proposed in
2002~\cite{abouraddy02}, and experimentally demonstrated as a
quantum-imaging technique in 2003 and 2004~\cite{nasr03,nasr04}.
Three-dimensional QOCT images of a biological sample, an onion-skin
tissue coated with gold nanoparticles, were obtained more recently
\cite{nasr09}, confirming that QOCT is a viable biological imaging
technique. This also represented the first demonstration that a
quantum-entangled entity could interact with a nonplanar,
scattering, biological medium to generate a collection of quantum
interferograms, which in turn could be used to construct a
biological image.

A particular merit of QOCT, as indicated above, is that it is
inherently immune to even-order group-velocity dispersion
(GVD)~\cite[chap.~22]{saleh07} imparted by the sample, by virtue of
the frequency entanglement associated with the twin-photon
pairs~\cite{franson92,steinberg92a,steinberg92b,larchuk95}.
Conventional OCT, in contrast, is subject to GVD, which degrades
resolution. Moreover, for sources of the same bandwidth, the
twin-photon entanglement endows QOCT with a resolution enhancement
of a factor of two in comparison with that achievable with
OCT~\cite{abouraddy02}.

In addition, QOCT permits a direct determination of the GVD
coefficients of the interstitial media between the reflecting
surfaces of the sample~\cite{nasr04}. A typical QOCT scan comprises
two classes of features. Features in the first class carry the
information that is most often sought in OCT: the depth and
reflectance of the layer surfaces that constitute the sample. In
QOCT, each of these features is associated with a reflection from a
single surface and is \emph{immune} to GVD. Features in the second
class arise from cross-interference between the reflection
amplitudes associated with every pair of surfaces and \emph{are
sensitive} to the dispersion characteristics of the media between
them. Measurement of the broadening of a feature associated with two
consecutive surfaces directly yields the GVD coefficient of the
interstitial medium lying between them. In an OCT scan, in contrast,
only a single class of features is observed; each feature is
associated with the reflection from a single surface and is subject
to the cumulative dispersion of the portion of the sample lying
above it. Thus, GVD information is not directly accessible via OCT;
to measure the GVD of a particular buried medium, it is necessary to
consecutively compute the GVD of each of the constituent layers
above it.

The performance characteristics of a QOCT system, such as its
signal-to-noise ratio (SNR) and speed, are determined by a number of
factors, including the optical power (biphoton flux) in the
interferometer \cite{nasr09}.

\paragraph{Quantum-mimetic optical coherence tomography (QM-OCT).}
QOCT has inspired a number of ``quantum-mimetic" versions that
operate on the basis of classical nonlinear optics
\cite{erkmen06,legouet10,resch07,kaltenbaek08}. Though these
approaches mimic the dispersion-cancellation feature inherent in
photon pairs that are entangled in frequency, they have their own
limitations, which are principally associated with the complexity of
the required experimental arrangements.

\paragraph{Polarization-sensitive quantum optical coherence tomography (PS-QOCT).}
As an extension of the QOCT theory proposed in 2002
\cite{abouraddy02}, the theory for Type-II polarization-sensitive
quantum optical coherence tomography (PS-QOCT) was set forth in
2004; it offered optical sectioning with polarization-sensitive
capabilities \cite{booth04a}. This approach provides a means for
obtaining information about the optical path length between
isotropic reflecting surfaces, the relative magnitudes of the
reflectances from each interface, the birefringence of the
interstitial materials, and the orientation of the optical axes of
the sample. As with its precursor, PS-QOCT is immune to even-order
sample dispersion and thus permits measurements to be carried out at
depths greater than those accessible via PS-OCT. A general
Jones-matrix theory for analyzing PS-QOCT systems, along with a
proposed procedure for carrying out these experiments, is provided
in Ref.~\cite{booth04a}.

\section{PS-QOCT Experiments} \label{sec:exp}

In this paper we report the results of several proof-of-principle
experiments illustrating that Type-II polarization-sensitive QOCT
provides results in accord with the theory \cite{booth04a}. The
experimental arrangement used in these experiments is discussed in
detail in Sec.~\ref{ssec:exparrange}.

In the first experiment (Sec.~\ref{ssec:GVDpsec}), we make use of a
reflective sample buried under a highly dispersive material to
demonstrate that even-order group-velocity dispersion cancellation
can be achieved when the pulsed pump laser is operated in the psec
regime. This results from the frequency anti-correlation of the
signal and idler downconverted photons.

The second experiment (Sec.~\ref{ssec:GVDfsec}), conducted with pump
pulses of fsec duration, illustrates the persistence of
group-velocity dispersion when the pump pulses have insufficient
duration. This is expected because the broad spectra of the pump
pulses blur the frequency anti-correlation of the signal and idler
photons. These experiments reveal that dispersion cancellation can
be achieved with sufficiently long pump pulses, or with CW.

In the third experiment (Sec.~\ref{ssec:nonparallel}), we show that
PS-QOCT can be successfully used for samples that scatter light
rather than reflect it, such as biological specimens, by inserting a
single lens at a carefully selected location in the experimental
apparatus. This experiment also demonstrates that high-resolution
lateral, as well as axial, imaging can be attained with PS-QOCT.

Finally, the fourth experiment (Sec.~\ref{ssec:twolayer}), conducted
with a sample comprising two surfaces and an interstitial medium,
reveals the presence of two classes of features in the Type-II QOCT
interferogram. As with Type-I QOCT, these correspond to interference
arising from each surface individually and to cross-interference
involving both surfaces. This experiment demonstrates that PS-QOCT
can be used to determine the GVD coefficients of the interstitial
media between the reflecting surfaces of the sample.

\subsection{Experimental arrangement}\label{ssec:exparrange}

\begin{figure}[htbp]
\centering\includegraphics[width=11.3cm]{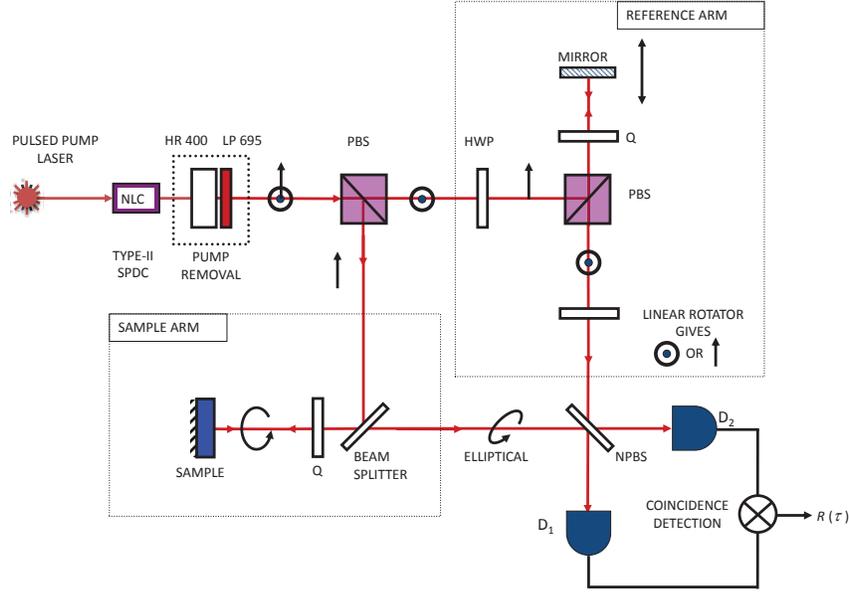}
\caption{Principal features of an experimental arrangement suitable
for implementing Type-II polarization-sensitive quantum-optical
coherence tomography (PS-QOCT) with a pulsed pump laser.}
\label{fig:QOCT-setup}
\end{figure}
A general experimental arrangement that can be used for
investigating the properties of pulse-pumped Type-II QOCT
experiments is displayed in Fig.~\ref{fig:QOCT-setup}. A
frequency-doubled Ti:sapphire laser, used to generate either 10-psec
or 80-fsec optical pulses at a center wavelength of 400~nm, pumps a
$\beta$-barium borate (BBO) nonlinear optical crystal (NLC) oriented
for Type-II, collinear, degenerate spontaneous parametric
downconversion (SPDC) at a center wavelength of 800~nm
\cite{booth04b}. The pump beam is removed from the SDPC by a highly
reflective mirror (HR~400) centered at the pump wavelength in
conjunction with a long-pass filter (LP~695).

The vertically polarized signal photons and horizontally polarized
idler photons are separated by a polarizing beam splitter (PBS) and
fed to the sample and reference arms of a Mach--Zehnder
interferometer, respectively. For the experiments considered here,
the polarization in the reference arm was set to horizontal
(indicated by $\odot$). The reference arm consists of a variable
path-length delay comprised of a half-wave plate (HWP), a polarizing
beam splitter (PBS), a quarter-wave plate (Q), and a translational
mirror that changes the length of the reference arm relative to the
sample arm. The final polarization of the reference beam can be
oriented to either vertical or horizontal by a linear rotator placed
before the final non-polarizing beam splitter (NPBS) of the
interferometer.

The sample arm contains a beam splitter, a quarter-wave plate (Q),
and the sample under study; this arrangement allows circularly
polarized light to impinge normally on the sample. The
back-scattered light from the sample, which in general has
elliptical polarization, is mixed with the delayed reference beam at
the final NPBS. In the general configuration portrayed in
Fig.~\ref{fig:QOCT-setup}, the elliptically polarized backscattered
light from the sample travels back along the probe-beam path,
through the NPBS in the sample arm, to the final non-polarizing beam
splitter.

The two outputs of the final NBPS are directed through pinholes (not
shown) to two single-photon counting detectors. The coincidence rate
$R(\tau)$ for photons arriving at the two detectors, as a function
of the path-length delay $c\tau$ in the reference arm, are recorded
in a time window determined by a coincidence-counting detection
circuit (indicated by $\otimes$).

\begin{figure}[htbp]
\centering\includegraphics[width=11.3cm]{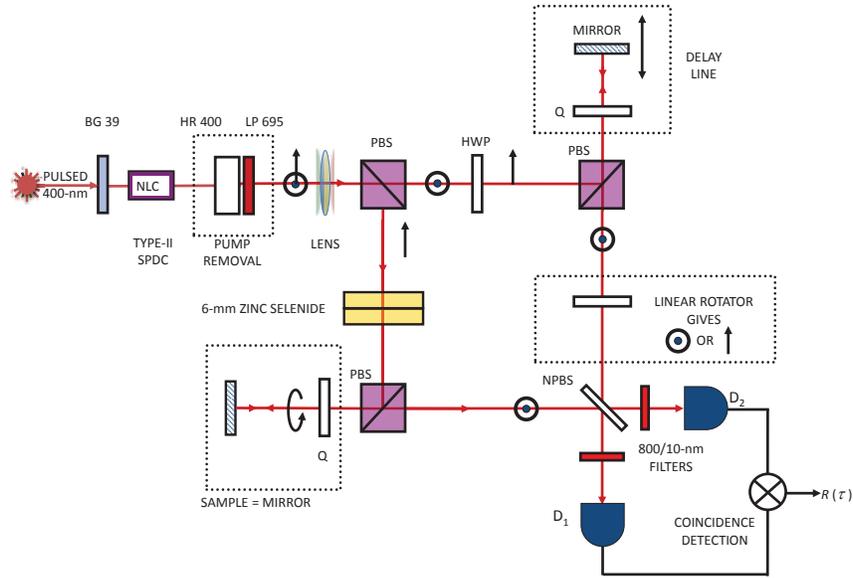}
\caption{Details of the setup used to carry out Type-II PS-QOCT
experiments with a pulsed pump laser. Details are provided in the
text.}\label{fig:type-II-QOCT}
\end{figure}
\begin{figure}[htbp]
\centering\includegraphics[width=6.5cm]{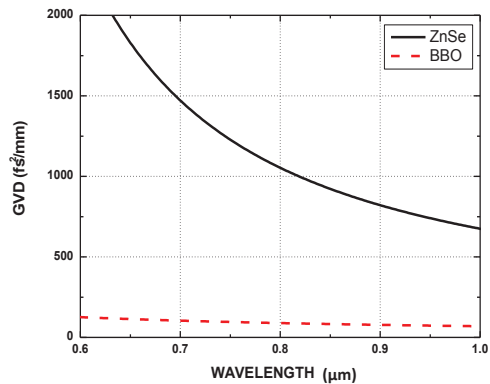}
\caption{Group-velocity dispersion parameter (fs$^2$/mm) versus
wavelength for zinc selenide (ZnSe, solid curve) and for
$\beta$-barium borate (BBO, dashed curve). At a wavelength of
800~nm, the GVD coefficient for ZnSe is more than 10 times greater
than that for BBO.} \label{fig:GVD-curves}
\end{figure}

The actual experimental setup used to carry out the experiments we
report is displayed in Fig.~\ref{fig:type-II-QOCT}. In experiments
designed to examine the dispersion-cancellation capabilities of the
pulse-pumped Type-II QOCT system, a reflective mirror sample was
buried beneath a 6-mm thickness of zinc selenide (ZnSe). At a
wavelength of 800~nm, this material is highly dispersive; its GVD
coefficient is more than 10 times greater than that of the BBO
nonlinear crystal, as is evident from Fig.~\ref{fig:GVD-curves}.

The light backscattered from the sample is directed through a
polarizing beam splitter in the sample arm, rather than through the
ordinary beam splitter depicted in Fig.~\ref{fig:QOCT-setup}. In
experiments designed to demonstrate that PS-QOCT can be used for the
imaging of scattered light, a lens is inserted before the initial
PBS.

\subsection{Cancellation of group-velocity dispersion for a ps-pulsed
pump}\label{ssec:GVDpsec}
\begin{figure}[t]
\centering\includegraphics[width=11.3cm]{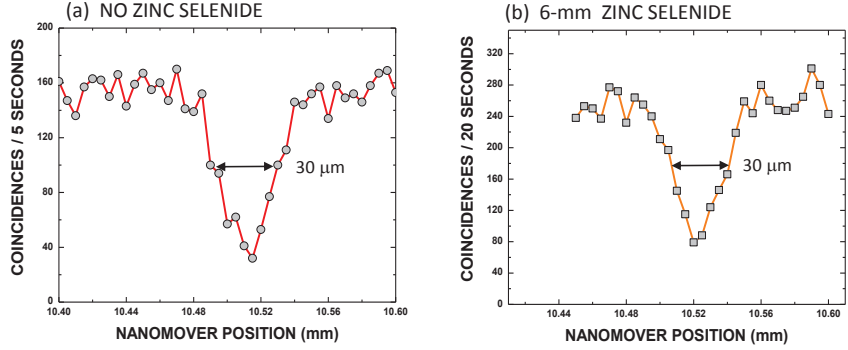}
\caption{Type-II QOCT coincidence rates $R(\tau)$ vs. path-length
delay with a ps-pulsed pump when the sample arm contains: (a) no
ZnSe, and (b) 6~mm of ZnSe. The solid curves connect the data points
to guide the eye. The measured 30-$\mu$m full-width-half-maximum
(FWHM) in (a), indicated by double arrows, is not broadened by the
presence of ZnSe in (b), indicating dispersion cancellation for
ps-pulsed pumping.} \label{fig:ps-disp}
\end{figure}
Using the setup portrayed in Fig.~\ref{fig:type-II-QOCT}, but with
no lens, we carried out a set of experiments in which the
Ti:sapphire laser was configured for psec-pulse operation. For such
long optical pulses, the source is effectively quasi-CW so that
even-order dispersion cancellation is expected.

The interferograms displayed in Fig.~\ref{fig:ps-disp} were obtained
with a 1.5-mm-thick BBO nonlinear crystal. Coincidence data are
shown: (a) in the absence of ZnSe, and (b) with 6~mm of ZnSe placed
in the sample arm before the single reflecting surface. Each data
point represents the number of coincidences measured in a 5-sec or
20-sec integration time (as indicated in Figs.~\ref{fig:ps-disp}(a)
and \ref{fig:ps-disp}(b), respectively), as the path-length delay
between the sample and reference arms was scanned via a
translational mirror in the reference arm located on a motorized
stage (Melles Griot Nanomover).

As the optical path-length delay approaches zero in
Fig.~\ref{fig:ps-disp}(a) (at nanomover position $\sim10.52$~mm),
quantum destructive interference at the final beam splitter causes
the observed coincidence rate to decrease, and the result is a sharp
dip in the $R(\tau)$ curve. This is the well-known Hong--Ou--Mandel
(HOM) interference dip \cite{hong87}. The interferogram in
Fig.~\ref{fig:ps-disp}(b) behaves similarly. Indeed, the 30-$\mu$m
full-width half-maximum (FWHM) of the dip displayed in
Fig.~\ref{fig:ps-disp}(b), in the presence of 6 mm of ZnSe, is the
same as that shown in Fig.~\ref{fig:ps-disp}(a) in the absence of
the ZnSe. This reveals that the presence of highly dispersive ZnSe
has no effect on the width of the dip, thereby demonstrating the
cancellation of group-velocity dispersion for ps-pulsed pumping.

\subsection{Persistence of group-velocity dispersion for a fs-pulsed pump}\label{ssec:GVDfsec}

The dispersion cancellation offered by QOCT is related to the
matching of each spectral component above the center frequency in
one beam to an anti-correlated spectral component below the center
frequency in its companion beam. For a CW pump beam with a single
frequency $\omega_{\rm p}$, energy conservation requires that
$\hslash\omega_{\rm p} = \hslash\omega_{\rm s} + \hslash\omega_{\rm
i}$, where the subscripts (p, s, i) denote the pump, signal, and
idler beams, respectively. Since the down-converted signal and idler
photon frequencies always sum to the pump frequency $\omega_{\rm
p}$, the signal and idler photons are perfectly anti-correlated in
frequency. The deviations from the center frequency ultimately
cancel via the interference of both beams at the final beam
splitter.

For a pulsed-pump beam, the same energy-conservation condition
applies. However, the range of frequencies comprising the pump pulse
increases as its duration decreases by virtue of their
Fourier-transform relationship. For a sufficiently short pump pulse,
therefore, the signal and idler frequencies need not sum to a
constant value, so that the signal and idler photons are no longer
precisely anti-correlated~\cite{grice97}. Consequently, the immunity
of QOCT to even-order dispersion in the sample is expected to
degrade as the pump-pulse duration shortens. It is clear from the
results reported in Sec.~\ref{ssec:GVDpsec} that 10-psec pump pulses
are sufficiently long so that dispersion cancellation is maintained.
However, the 80-fsec pump pulses are sufficiently short so that
dispersion cancellation is expected to fail.

To confirm this, we carried out a set of experiments similar to
those reported in Sec.~\ref{ssec:GVDpsec} but with the Ti:sapphire
laser reconfigured to generate 80-fs, rather than 10-ps, optical
pulses. Decreasing the pulse duration causes the peak pulse
intensity at the output of the Ti:sapphire laser to increase
substantially, thereby increasing the second-harmonic output power
fed to the nonlinear crystal. As a result, these experiments were
conducted using a thinner (0.5-mm thickness) BBO crystal. The
measured spectral bandwidth of the pump laser at its 400-nm center
wavelength was found to be $\Delta\lambda \approx 2$~nm.
\begin{figure}[t]
\centering\includegraphics[width=11.3cm]{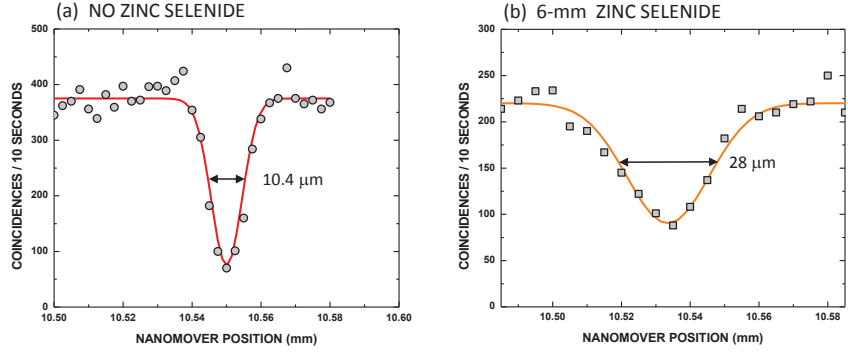}
\caption{Type-II QOCT coincidence rates $R(\tau)$ vs. path-length
delay with a fs-pulsed pump when the sample arm contains: (a) no
ZnSe, and (b) 6~mm of ZnSe. The solid curves represent Gaussian fits
to the quantum-interference dips. The measured FWHM in (b) is
substantially greater than that in (a), indicating the persistence
of group-velocity dispersion for fs-pulsed pumping.}
\label{fig:fs-disp}
\end{figure}

Coincidence data are displayed in Fig.~\ref{fig:fs-disp}(a) in the
absence of ZnSe, and in Fig.~\ref{fig:fs-disp}(b) in the presence of
6~mm of ZnSe placed in the sample arm before the single reflecting
surface. Each data point represents the number of coincidences
measured during a 10-sec integration time, as the path-length
difference between the sample and reference arms was scanned. Again,
dips in the $R(\tau)$ curves are observed as the optical path-length
delay approaches zero (at nanomover position $\sim10.53$~mm). The
solid curves represent Gaussian fits to the HOM quantum-interference
dips and the measured FWHM values are indicated by the double
arrows. In the absence of ZnSe, the width of the dip is 10.4~$\mu$m,
whereas in its presence the width is broadened to 28~$\mu$m. This
clearly demonstrates the persistence of group-velocity dispersion
when using a fs-pulsed pump to generate the SPDC. These results are
in sharp contrast to those obtained when using a ps-pulsed pump, as
portrayed in Fig.~\ref{fig:ps-disp}.

\subsection{Imaging of a scattering sample}\label{ssec:nonparallel}

QOCT is an interferometric technique that relies on the interference
between two spatial/polarization modes at the final beam splitter.
If the overlap of these modes is less than ideal, the quality of the
observed interference pattern will be degraded and the contrast of
the interference pattern diminished. The experiments reported in
Secs.~\ref{ssec:GVDpsec} and \ref{ssec:GVDfsec} made use of a mirror
sample that directly reflected the normally incident probe-beam
light so that it achieved spatial/polarization matching.

However, the back-scattered light returned from real samples often
extends over a cone of angles. The question arises as to whether a
lens inserted in the apparatus can serve to suitably collect this
returned light and thereby to maintain the integrity of the PS-QOCT
interferogram. Since a scattering sample, such as a biological
specimen, can be modeled as a collection of small reflecting regions
at various angles of tilt, we have carried out experiments using a
reflecting-mirror sample tilted over a range of angles to
demonstrate that Type-II PS-QOCT is suitable for imaging scattering
specimens, much as has been found with Type-I QOCT \cite{nasr09}.

\begin{figure}[t]
\centering\includegraphics[width=7cm]{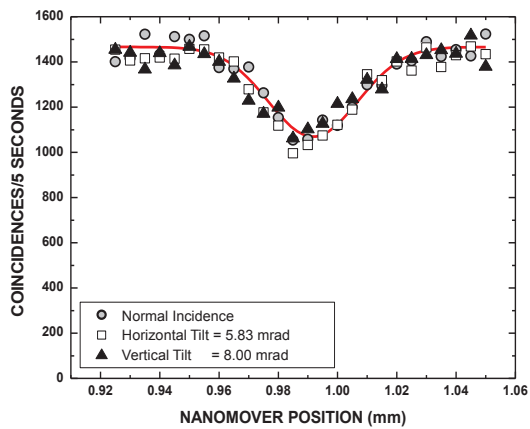} \caption{Type-II
QOCT interferograms obtained from a reflector tilted at various
angles to normal incidence. The experiments were conducted using the
experimental arrangement shown in Fig.~\ref{fig:type-II-QOCT}
configured for psec-pulse operation; the ZnSe was removed and a
3-cm-focal-length lens was placed before the mirror sample. Data are
presented for no mirror tilt (circles), 5.83-mrad horizontal mirror
tilt (squares), and 8.00-mrad vertical mirror tilt (triangles). The
visibility of the interferogram is not diminished by mirror tilt,
either horizontal or vertical. The solid curve represents a Gaussian
fit to the data for normal incidence.} \label{fig:lens-data}
\end{figure}

The first set of experiments was conducted using the experimental
setup displayed in Fig.~\ref{fig:type-II-QOCT} with a
3-cm-focal-length lens inserted before the mirror sample.
Figure~\ref{fig:lens-data} displays the PS-QOCT interferograms for:
1) reflection of the probe beam at normal incidence (circles); 2)
reflection at a 5.83-mrad mirror tilt in the horizontal direction
(squares); and 3) reflection at an 8.00-mrad mirror tilt in the
vertical direction (triangles). The data reveal that the sample
mirror could be tilted by more than 5 mrad in both the horizontal
and vertical directions without degrading the quantum-interference
pattern.

It is apparent from the results presented in
Fig.~\ref{fig:lens-data}, however, that the introduction of the lens
reduces the visibilities of all three interference patterns by
$\sim$25\%. It was ascertained that this results from an imbalance
in the interferometer that arises from a mismatch in the
spatial-mode overlap at the final beam splitter. This in turn leads
to distinguishability between the paths and a concomitant reduction
in the interferogram visibility.

\begin{figure}[t]
\centering\includegraphics[width=7cm]{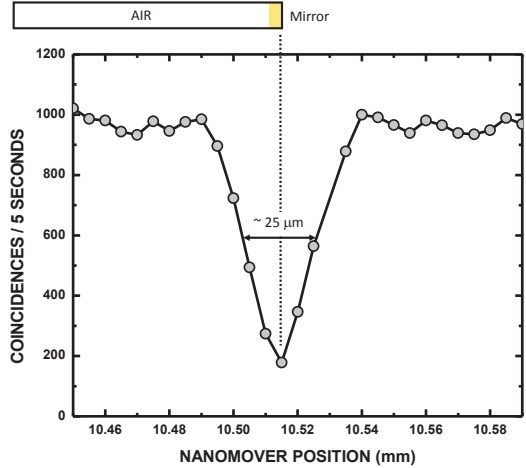}
\caption{Type-II QOCT interferogram obtained from a single reflector
at normal incidence. The experiments were conducted using the
experimental arrangement shown in Fig.~\ref{fig:type-II-QOCT},
configured for psec-pulse operation and no ZnSe; a single
20-cm-focal-length lens was placed before the initial polarizing
beam splitter (see Fig.~\ref{fig:type-II-QOCT}). The observed
visibility is $\sim$85\%. This interferogram represents the current
state-of-the-art in Type-II QOCT system performance.}
\label{fig:ps-pump-mirror}
\end{figure}

We therefore conducted a second set of experiments in which we
demonstrated that high-visibility interference can in fact be
regained by inserting an identical lens in the reference arm to
restore indistinguishability. Alternatively, we can place a single
lens before the first polarizing beam splitter (as shown in
Fig.~\ref{fig:type-II-QOCT}), since this symmetrically affects both
arms of the interferometer. In Fig.~\ref{fig:ps-pump-mirror}, we
display the interferogram obtained from a third set of experiments
in which we employed this latter technique. The spatial modes in the
sample and reference arms are modified in an identical manner, and
this does indeed lead to a substantially increased visibility of
$\sim$85\%, with a coincidence count rate of $\sim200$
coincidences/second. While the presence of this particular lens
restricts the sample tilt to $\le$~0.5~mrad, it increases the
overall count rate via improved mode matching at the fiber-coupled
single-photon counting detectors. These results can, no doubt, be
further improved by optimally selecting matched focusing systems for
the sample and reference arms.

Moreover, since the lens restricts the light falling on the sample
to a small lateral spot, these results demonstrate that
high-resolution lateral (transverse), as well as axial, imaging can
be attained for a PS-QOCT imaging system \cite{nasr02}. The lateral
resolution can be enhanced by making use of high-numerical-aperture
(NA) focusing optics in the sample arm. This approach should be
implemented carefully, however, since incorporating a single lens in
the apparatus can impair axial resolution, as discussed above.
Rather, matched focusing in both arms of the interferometer should
be used to preserve indistinguishability. The construction of a
two-dimensional imaging system of this kind would benefit from the
combined and coordinated scanning of the focusing lens in the sample
arm and the interferometric delay. This can be achieved by using a
single mechanical stage to control both functions, as has been
successfully implemented in OCT \cite{izatt94}.

\subsection{Determination of interstitial-media dispersion}\label{ssec:twolayer}

The final experiment that we report is designed to demonstrate that
PS-QOCT can serve as a tool for determining the GVD coefficients of
the interstitial (in general birefringent) media between the
reflecting interfaces of a multi-layer sample.

For a sample with two reflective surfaces, PS-QOCT theory predicts
two classes of features in the Type-II QOCT interferogram
\cite[Sec.~IVC]{booth04a}. These correspond, respectively, to a pair
of interference dips, each associated with one of the surfaces, and
to a cross-interference feature involving both surfaces that is
expected to appear midway between the two dips
\cite[Eq.~(43)]{booth04a}. This latter feature can be either a peak
or a dip, depending on the values of the pump frequency $\omega_{\rm
p}$, the average refractive index ${\bar n}$, the nonlinear crystal
thickness, and the arguments of the complex reflection coefficients
$r_0$ and $r_1$. A numerical simulation for the interferogram
expected for a quartz flat is instructive \cite[Fig.~5]{booth04a}.

\begin{figure}[t]
\centering\includegraphics[width=7cm]{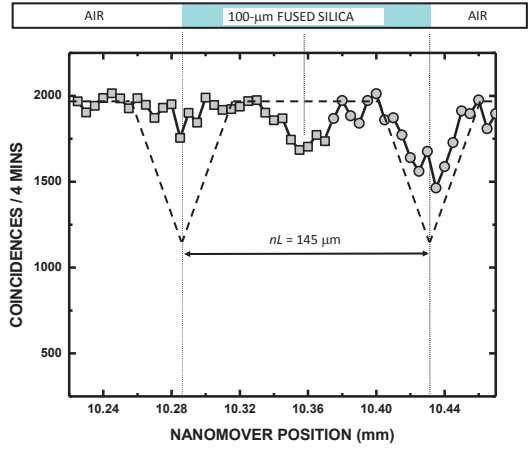}
\caption{Type-II QOCT interferogram obtained from a silica flat of
refractive index $n=1.45$ and thickness $L = 100\:\mu$m. The
experiment was conducted using the setup shown in
Fig.~\ref{fig:type-II-QOCT}, configured for psec-pulse operation and
no ZnSe; a 20-cm-focal-length lens was placed before the initial
polarizing beam splitter. The data points indicated by squares
represent coincidence count rates acquired over 4 minutes, whereas
the circles represent renormalized coincidence count rates acquired
over 1 minute. As expected, the two interference dips are separated
by the optical path length of the sample $nL = 145\:\mu$m; however,
they exhibit asymmetric visibilities as a result of alignment
errors. The dashed curve is the theoretical prediction for the two
interference dips.} \label{fig:ps-pump-silica}
\end{figure}

Our experiment was carried out using the apparatus portrayed in
Fig.~\ref{fig:type-II-QOCT}, with a 20-cm-focal-length lens inserted
before the first polarizing beam splitter. The sample was a silica
flat of refractive index $n=1.45$ and thickness $L = 100\:\mu$m.
From the PS-QOCT theory discussed above, we expect two dips in the
interferogram associated with the reflections at the two surfaces of
the flat. These dips should have equal visibility for this thin
sample since the intensity reflectances from the two interfaces will
be equal at normal incidence: $|r_0|^2 = |r_1|^2 = [(n-1)/(n+1)]^2 =
0.034$. The dips should be separated by the optical path length of
the sample, $nL = 145\:\mu$m. The interference pattern expected from
the two interfaces is illustrated as the dashed curve in
Fig.~\ref{fig:ps-pump-silica}.

The observed QOCT interferogram for the silica flat is also
displayed in Fig.~\ref{fig:ps-pump-silica}. The data indicated by
squares represent coincidence count rates acquired over 4 minutes,
whereas the circles represent renormalized coincidence count rates
acquired over 1 minute. The interferogram is constructed from two
separate scans because of operational problems with the Ti:sapphire
laser that prevented all of the data from being acquired in a single
scan. Though the experiment was not optimized, the interference dips
resulting from the two interfaces are clearly evident. They are at
the correct locations and are indeed separated by the optical path
length of the sample, $nL = 145\:\mu$m, although they exhibit
different visibilities because of alignment problems.

Also, as expected, the cross-interference feature emerges midway
between the dips. Its width, in conjunction with those of the two
interference dips, can be used to provide the GVD coefficients for
the interstitial silica.

For multilayer samples it is worth emphasizing that the widths of
the cross-interference terms are determined only by the dispersion
of the medium residing between the two relevant reflective surfaces,
and not by the nature of the material under which they are buried.
Furthermore, in the absence of prior information relating to the
structure of the sample, interference features in the PS-QOCT
interferogram associated with reflections from individual surfaces
may be confounded by the presence of cross-interference features
associated with pairs of surfaces. These two classes of features may
be readily distinguished, however, since slight variations of the
pump frequency can be used to change the form of features in the
second class (e.g., from dips to humps), whereas those in the first
class are invariant to such variations. Thus, dithering the pump
frequency, for example, can serve to wash out the features in the
second class, thereby leaving intact the desired
dispersion-cancelled portion of the interferogram that reveals the
axial structure of the sample \cite{nasr04}. We expect that returns
from scattering media would exhibit similar behavior because of the
randomness of the relative phases associated with different
surfaces. At the same time, simple subtraction of this pattern from
the undithered version would allow the second class of features to
be separated, thereby facilitating the determination of the GVD
coefficients of the various media comprising the sample.

\begin{figure}[t]
\centering\includegraphics[width=9cm]{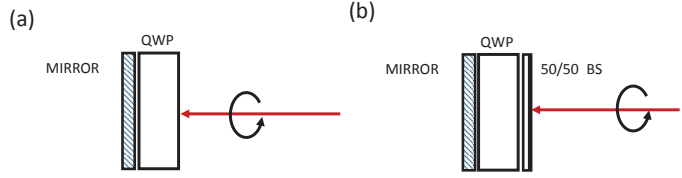}
\caption{Samples expected to be useful for extended studies
examining other properties of a PS-QOCT system. The single- and
dual-layer samples are, respectively, (a) a concatenation of a
quartz plate and a highly reflective mirror; and (b) a 50/50 beam
splitter integrated with the sample specified in (a). The probe
light is incident from the right.} \label{fig:proposed-samples}
\end{figure}

The experiment discussed above can be extended to provide a more
thorough examination of the polarization-sensitive properties of the
PS-QOCT paradigm. For example, an experiment could be carried out by
replacing the polarizing beam splitter in the sample arm of
Fig.~\ref{fig:type-II-QOCT} by an ordinary beam splitter (as shown
in Fig.~\ref{fig:QOCT-setup}), so that the full polarization
information inherent in the sample returns along the sample path to
the final non-polarizing beam splitter. Two idealized samples,
illustrated in Figs.~\ref{fig:proposed-samples}(a) and (b), could
serve to establish the validity of Figs.~4 and~5
of~Ref.~\cite{booth04a}, respectively. The highly reflective mirror
in Fig.~\ref{fig:proposed-samples}(a), as well as the 50/50 beam
splitter and mirror in Fig.~\ref{fig:proposed-samples}(b), offer far
stronger reflections than those from the surfaces of the quartz
plate (for which $|r|^2 = 0.034$), which would facilitate this
experiment by shortening the integration time required for a scan
such as that shown in Fig.~\ref{fig:ps-pump-silica}.

The use of samples such as these would readily permit a set of three
measurements to yield: 1) the thickness of the quartz; 2) its
birefringence properties; and 3) its orientation angle. This is
because it has previously been established that three experiments
are required to completely determine the properties of the sample
\cite{booth04a}. The procedure is as follows: The reference-arm
polarization is first selected to be horizontal (H) and the
associated quantum interferogram $R_{\rm H}$ is measured by
recording the photon coincidence rate from the two detectors as the
path-length delay $c\tau$ is scanned. The second experiment is
performed by rotating the reference-arm polarization to the
orthogonal vertical (V) position and measuring the quantum
interferogram $R_{\rm V}$. The third experiment is carried out by
selecting a value of $c\tau$ that coincides with the position of a
layer; the angles of the polarization elements in the reference arm
are then adjusted to maximize the coincidence rate.

The sample properties are obtained via the following procedure
\cite{booth04a}: The birefringence inherent in the parameter
$\delta$ is determined by forming the ratio of $\Lambda_{\rm V}$ and
$\Lambda_{\rm H}$ at a value of $c\tau$ that coincides with the
location of a layer. The quantity $\alpha$ is determined from the
angles of the polarization elements in the reference arm and by
solving the equations for orthonormality. This approach is not
unlike the nulling techniques used in ellipsometry; the total
quantum interferogram $R_{\rm T}$ is computed from the sum of
$R_{\rm H}$ and $R_{\rm V}$, and is then readjusted for the DC
offset given by the constant term $\Lambda_0$, so that $R_{\rm T} =
(R_{\rm V} + R_{\rm H} - \Lambda_0)/\Lambda_0$. The $R_{\rm T}$
curve provides the path-length delay between the interfaces as well
as the ratio of the relative reflectances from each layer.

\section{Discussion}\label{sec:discussion}

\paragraph{Signal-to-noise ratio and speed.}
The signal-to-noise ratio (SNR) and speed of the PS-QOCT technique
are determined by a number of factors, including the optical power
(biphoton flux) in the interferometer, the transmittances of the
optical paths, the quantum efficiencies of the detectors, and the
duration of the experiments. Although the biphoton flux can be
increased by simply raising the pump power, an upper limit is
imposed by the saturation level of the single-photon detectors and
coincidence circuit. Faster single-photon detectors, such as those
relying on superconducting technology~\cite{marsili08}, as well as
faster coincidence circuits, should significantly enhance the SNR
and speed of PS-QOCT. Additionally, some of the recent advances in
biphoton generation that have enabled ultrahigh axial resolution to
be achieved in Type-I QOCT \cite{nasr08} can be considered in the
context of Type-II QOCT. Also in the offing are electrically driven
solid-state biphoton sources~\cite{hayat08} that promise optical
powers in the $\mu$W region, which could enhance the prospects for
PS-QOCT.

\paragraph{Generalization of the model.}
To advance the use of PS-QOCT, it would also be beneficial to relax
some of the simplifying assumptions used in the initial model
\cite{booth04a}. The specimen can be endowed with properties that
more closely mirror biological tissue rather than being considered
as a collection of layers separated by transparent dispersive media.
It would be helpful to accommodate: 1) Depletion of the probe beam
--- the assumption of an undepleted probe beam is widely used
throughout the OCT literature but the ability of PS-QOCT to probe
more deeply into tissue may make depletion more important; 2)
Frequency- and polarization-dependent reflection coefficients at the
boundaries; 3) Scattering and absorption at the boundaries and in
the interstitial layers.

\section{Conclusion}\label{sec:concl}
We have carried out a collection of experiments that have served to
demonstrate the operation of dispersion-cancelled and
dispersion-sensitive Type-II polarization-sensitive quantum optical
coherence tomography (PS-QOCT). We have illustrated the principal
advantages that stem from the entanglement of the twin-photon
source: dispersion cancellation, polarization sensitivity, and the
ability to directly measure the dispersion characteristics of the
interstitial media. We have also shown that the judicious placement
of focusing elements in the experimental apparatus allows
high-resolution lateral, as well as axial, imaging to be achieved.
Suggestions have been provided for further experiments.

\section*{Acknowledgments} This work was supported by a U.S. Army
Research Office (ARO) Multidisciplinary University Research
Initiative (MURI) Grant; and by the Bernard M. Gordon Center for
Subsurface Sensing and Imaging Systems (CenSSIS), an NSF Engineering
Research Center.


\end{document}